\documentstyle [12pt,openbib]{article}

\font\tenmsb=msbm10  scaled \magstep1
\font\sevenmsb=msbm7 scaled \magstep1
\font\fivemsb=msbm5  scaled \magstep1
\newfam\msbfam
\textfont\msbfam=\tenmsb  \scriptfont\msbfam=\sevenmsb
  \scriptscriptfont\msbfam=\fivemsb
\def\hexnumber@#1{\ifcase#1 0\or1\or2\or3\or4\or5\or6\or7\or8\or9\or
      A\or B\or C\or D\or E\or F\fi }
\edef\msb@{\hexnumber@\msbfam}
\mathchardef\subsetneq="3\msb@28

\begin{document}

\vspace{-5.5ex}
\rightline{solv-int/9310001}

\vspace{1ex}
{\noindent\Large\bf Deformations of Calogero-Moser Systems}\footnote{Talk
given at the 9th Workshop on {\sl Nonlinear Evolution Equations and Dynamical
Systems \\  (NEEDS '93)}, held at Gallipoli, Italy, September 3-12, 1993.}

\vspace{3ex}
{\noindent\large J.F. van Diejen}\footnote{E-mail
address: jand@fwi.uva.nl}

{\noindent\small
Department of Mathematics and Computer Science, University of Amsterdam,\\
Plantage Muidergracht 24, 1018 TV Amsterdam, The Netherlands.}

\vspace{1ex}
{\small\noindent {\bf Abstract.}
Recent results are surveyed pertaining to the complete integrability of
some novel $n$-particle models in dimension one.
These models generalize the Calogero-Moser systems
related to classical root systems.}

\vspace{3ex}
\noindent {\bf 1. Introduction} \hfill

\noindent The Hamiltonian of the celebrated Calogero-Moser (CM) system
\cite{op}
is given by
\begin{equation}\label{cm}
H_{cm}= 1/2 \sum_{1\leq j\leq n} \theta_j^2 +
g^2 \sum_{1\leq j < k\leq n} \wp (x_j-x_k) ,
\end{equation}
where $\wp (z)$ denotes the Weierstra\ss{} $\wp$-function \cite{ww}
or a degeneration thereof ($1/z^2$, $1/{\rm sh}^2(z)$ or $1/\sin^2 (z)$).
The integrability of $H_{cm}$ (\ref{cm}) was proved with the aid of a
Lax matrix \cite{op}.

Some years ago a relativistic generalization of $H_{cm}$ was
introduced \cite{rs,r1,r2}. The Hamiltonian of the relativistic system
(RCM) reads
\begin{equation}\label{rcm}
H_{rcm}=\sum_{1\leq j\leq n}  {\rm ch} (\beta\theta_j)
\prod_{k\neq j} \left[ 1+ \beta^2g^2 \wp (x_j-x_k) \right]^{1/2} .
\end{equation}
One can look upon the RCM system as a one-parameter deformation of the
CM model, with $\beta \sim 1/c$ (the inverse of the speed of light)
acting as deformation parameter.
For $\beta \rightarrow 0$, which corresponds to the nonrelativistic
limit, $\beta^{-2}(H_{rcm}(\beta )-n)$ converges to $H_{cm}$.
The relativistic system is also integrable;
explicit formulas have been found for a complete set of
integrals in involution:
\begin{equation}
H_{l,\, rcm} =
\sum_{\stackrel{J\subset \{ 1,\ldots ,n\} }{|J|=l}}
e^{-\beta\sum_{j\in J}\theta_j}\,
\prod_{\stackrel{j\in J}{k\in J^c}}
\left[ 1+\beta^2g^2 \wp (x_j-x_k) \right]^{1/2},\;\;\;\;\;\;\;\;\;\;
l=1,\ldots ,n.
\end{equation}

{}From a Lie-theoretic perspective the above $n$-particle models are
connected with the root system $A_{n-1}$. Here, we will take a look at
similar deformations of the CM systems related to
classical root systems other than $A_{n-1}$ (i.e. $B_n$, $C_n$, $D_n$ and
$BC_n$).
A more detailed discussion of the material covered below
(including proofs) can be found in the papers \cite{die1,die2}.

\vspace{2ex}
\noindent {\bf 2. Trigonometric Potentials} \hfill

\noindent In the case of trigonometric potentials our system is characterized
by
the Hamiltonian
\begin{equation}
H = \sum_{1\leq j\leq n}\left( {\rm ch} (\beta \theta_j)\,
V_j^{1/2}V_{-j}^{1/2}\, -\,
(V_j+V_{-j})/2 \right) \label{Htr}
\end{equation}
with
\begin{eqnarray}\label{V1}
V_{\varepsilon j} &=& w(\varepsilon x_j)
\prod_{k\neq j} v(\varepsilon x_j+x_k) v(\varepsilon x_j-x_k),\;\;\;\;\;\;\;\;
\varepsilon =\pm 1, \\
v(z) &=& \frac{\sin\alpha (\mu +z)}{\sin (\alpha z)}, \\
w(z) &=& \frac{\sin\alpha (\mu_0 +z)}{\sin (\alpha z)}
         \frac{\cos\alpha (\mu_1 +z)}{\cos (\alpha z)}
         \frac{\sin\alpha (\mu_0^\prime  +z)}{\sin (\alpha  z)}
         \frac{\cos\alpha (\mu_1^\prime  +z)}{\cos (\alpha  z)}
                \label{trig} .
\end{eqnarray}
One can look upon the functions $v$ and $w$ as potentials:
$v$ governs the interaction between the particles and $w$ models an external
field.
The parameters $\mu$, $\mu_r$ and $\mu_r^\prime$ ($r=0,1$) act as
coupling constants; after setting them equal to zero the particles become
free ($v,w=1$).

Just as for the RCM system, explicit formulas have been found
that constitute a complete set of integrals in involution for the
Hamiltonian $H$~(\ref{Htr})-(\ref{trig})
\begin{equation}
H_l =
\sum_{\stackrel{J\subset \{ 1,\ldots ,n\} ,\, |J|\leq l}
               {\varepsilon_j=\pm 1,\, j\in J}          }
{\rm ch} (\beta \theta_{\varepsilon J})\,
V_{\varepsilon J;\, J^c}^{1/2}\, V_{-\varepsilon J;\, J^c}^{1/2}\, U_{J^c,\,
l-|J|},
\;\;\;\;\;\;\;\; l=1,\ldots ,n,
\end{equation}
with
\begin{eqnarray}
\theta_{\varepsilon J}&=& \sum_{j\in J}\; \varepsilon_j\, \theta_j, \\
V_{\varepsilon J;\, K} &=& \prod_{j\in J} w(\varepsilon_jx_j)\,
\prod_{\stackrel{j,j^\prime \in J}{j<j^\prime}}
v^2(\varepsilon_jx_j+\varepsilon_{j^\prime}x_{j^\prime})\,
\prod_{\stackrel{j\in J}{k\in K}} v(\varepsilon_j x_j+x_k)
v(\varepsilon_j x_j -x_k), \label{V} \\
U_{I,p}&=& \sum_{\stackrel{1\leq q\leq p}{\varepsilon_i=\pm 1,\, i\in I}}
(-1)^q
\sum_{
   \stackrel{\emptyset\subsetneq I_1\subsetneq \cdots \subsetneq I_q\subset I}
            {|I_q|=p}        } \;
\prod_{1\leq q^\prime \leq q}
V_{\varepsilon (I_{q^\prime} \setminus I_{q^\prime -1});\,
                I\setminus I_{q^\prime}}
\end{eqnarray}
($U_{I,0}=1$, $I_0=\emptyset$). Notice that $H_1$ coincides with the
Hamiltonian $H$~(\ref{Htr})
up to a factor two.

\vspace{1ex}
\noindent {\bf Theorem 1 (Liouville integrability):} {\sl The functions
$H_1,\ldots ,H_n$
are in involution (with respect to the standard Poisson bracket
induced by the symplectic form $\omega =\sum_j dx_j \wedge d\theta_j$).}
\vspace{1ex}

After reparametrization according to
\begin{equation}\label{parres}
\mu =i\beta g, \;\;\;\;\; \mu_r =i\beta g_r, \;\;\;\;\;
                     \mu_r^\prime =i\beta g_r^\prime ,
\end{equation}
the asymptotics for $\beta \rightarrow 0$ leads to the CM
system associated with the root system $BC_n$:
\begin{equation}
H_1 (\beta) = H_{1,0}\, \beta^{2}\;\; +\; o(\beta^{2}),
\end{equation}
with
\begin{eqnarray}
H_{1,0}&=&\sum_{1\leq j\leq n} \theta_j^2 +
g^2\alpha^2 \sum_{1\leq j\neq k\leq n}
        \left(\, \frac{1}{\sin^{2}\alpha (x_j+x_k)} +
        \frac{1}{\sin^{2}\alpha (x_j-x_k)}  \, \right)  \nonumber \\
& & \;\;\;\;\;\;\;\;\;\;\;\;\;\;\;\;\;\;\;\;\;\;\;\; +\alpha^2 \sum_{1\leq
j\leq n}
  \left( \, \frac{(g_0+g_0^\prime)^2}{\sin^{2}(\alpha x_j)} +
   \frac{(g_1+g_1^\prime)^2}{\cos^{2}(\alpha x_j)} \, \right)
   \;\; + \; const . \label{Htr0}
\end{eqnarray}
More generally, one has
\begin{equation}\label{expb}
H_l(\beta ) = H_{l,0}\, \beta^{2l} \;\; + \; o(\beta^{2l})
\end{equation}
with
\begin{equation}
H_{l,0} =
\sum_{\stackrel{J\subset \{1,\ldots ,n\} }{|J|=l}} \;
\prod_{j\in J} \theta_j^2 \;\; + l.o. \label{leadpdo}
\end{equation}
($l.o.$ stands for terms of lower order in the momenta $\theta_j$).
An immediate consequence of Theorem~1 and the above expansion formulas is
the integrability of the CM Hamiltonian $H_{1,0}$ (\ref{Htr0}).

\vspace{1ex}
\noindent {\bf Theorem 2 (transition to the $BC_n$-type CM system):} {\sl The
limits
\begin{equation}
H_{l,0} =\lim_{\beta \rightarrow 0}\; \beta^{-2l} H_l(\beta ),
\;\;\;\;\;\;\;\;\;\; l=1,\ldots ,n,
\end{equation}
exist and the resulting functions $H_{1,0},\ldots ,H_{n,0}$ are in involution.}

\vspace{2ex}
\noindent {\bf 3. Elliptic Potentials} \hfill

\noindent As elliptic counterpart of $H$ (\ref{Htr})-(\ref{trig}), I propose
the
following Hamiltonian:
\begin{equation}
H = \sum_{1\leq j\leq n} {\rm ch} (\beta \theta_j)\,
V_j^{1/2}V_{-j}^{1/2}\;\;\;
+\; U \label{Hel}
\end{equation}
where $V_{\varepsilon j}$ is again of the form (\ref{V1}),
but now with $v$ and $w$ given by
\begin{equation}
v(z) = \frac{\sigma (\mu +z)}{\sigma (z)}, \;\;\;\;\;\;\;\;\;\;\;\;\;\;\;
w(z) = \prod_{0\leq r\leq 3}
         \frac{\sigma_r(\mu_r +z)\; \sigma_r(\mu_r^\prime  +z)}
                    {\sigma_r(z)\; \sigma_r(z)}.\label{elw}
\end{equation}
The function $U$ is defined by
\begin{eqnarray}
U &=& \sum_{0\leq r\leq 3} c_r
  \prod_{1\leq j\leq n} v(\omega_r+ x_j) v(-\omega_r -x_j) ,\label{U} \\
c_r &=& \sigma (\mu )^{-2}
\prod_{0\leq s\leq 3} \sigma_s(\mu_{\pi_r(s)})
                      \sigma_s(\mu^\prime_{\pi_r(s)}), \label{cr}
\end{eqnarray}
where the following permutations have been introduced:
$\pi_0= id$, $\pi_1 =(01)(23)$, $\pi_2=(02)(13)$ and $\pi_3=(03)(12)$.
In the above expressions $\sigma (z)$ denotes the Weierstra\ss{}
$\sigma$-function
with quasi-periods $2\omega_r$, $r=1,2,3$, and the
$\sigma_r$ are the associated functions \cite{ww}
\begin{equation}
\sigma_r(z)= \exp ( -\eta_r z)\sigma (\omega_r +z)/\sigma (\omega_r),
\;\;\;\;\;\;\;\;\;\;\; r=1,2,3.
\end{equation}
(By convention $\omega_0 \equiv 0$ and $\sigma_0(z)\equiv \sigma(z)$).

Although the integrability of $H$ (\ref{Hel})-(\ref{cr}) has not been
demonstrated
yet, some partial results have been obtained \cite{die2}:
{\em i.} I found an additional Hamiltonian, which commutes with
$H$ (\ref{Hel})-(\ref{cr}) if the coupling constants of the external field
satisfy
the condition $\sum_{0\leq r\leq 3}\; (\mu_r+\mu_r^\prime)=0$;
{\em ii.} for special values of the coupling constants $H$
(\ref{Hel})-(\ref{cr})
can be seen as a reduction of the RCM Hamiltonian (\ref{rcm}); the
integrability of the
model then follows from \cite{r1};
{\em iii.} if $\mu $ equals a half-period $\omega_r$, then straightforward
generalization of the expressions in Section~2 results in an ansatz for
the higher integrals; their commutativity has been verified for $n\leq 4$.

After setting parameters as in Eq.~(\ref{parres}), one has
\begin{equation}
H(\beta ) = const\; +\; H_0\,\beta^2/2\;\;\; +o(\beta^2)
\end{equation}
with
\begin{equation}
H_0 = \sum_{1\leq j\leq n} {\theta}_j^2+\;\;
g^2 \sum_{1\leq j\neq k\leq n}
            \left( \wp (x_j+x_k)+\wp (x_j-x_k)\right) \nonumber\\
      \; +\sum_{\stackrel{1\leq j\leq n}{0\leq r\leq 3}}
    (g_r+g_r^\prime)^2\,\wp (\omega_r +x_j) . \label{Hel0}
\end{equation}
A Lax pair for the flow generated by $H_0$~(\ref{Hel0}) has been presented
by Inozemtsev \cite{ino}.

\vspace{2ex}
{\em Remarks: i.} For special values of the parameters $g_r$, $g_r^\prime$,
the Hamiltonians $H_{1,0}$~(\ref{Htr0}) and $H_0$~(\ref{Hel0})
reduce to CM Hamiltonians that are
associated with the root systems $B_n$, $C_n$ and $D_n$.

{\em ii.} Quantization of the Hamiltonians for $\beta \neq 0$ gives rise to
difference operators instead of the usual partial differential operators.
The reason is the occurrence of
exponentials of the form $\exp (\pm \beta i\partial_{x_j})$ in the
quantized Hamiltonian.
In the case of trigonometric potentials the eigenfunctions of the quantum
system
turn out to be the product of a factorized ground-state wave function and
recently discovered multivariable generalizations of the Askey-Wilson
polynomials
\cite{die1,die2}.

{\em iii.} Further limits of the Hamiltonian with elliptic potentials
in Eqs.~(\ref{Hel})-(\ref{cr}) lead to novel $n$-particle models with nearest
neighbor interaction \cite{die3}.
These models generalize the nonperiodic relativistic Toda chain
\cite{rt}, and form a deformation of known Toda chains with
very general boundary conditions \cite{ino2}.

\end{document}